\documentclass{aastex}

\newcommand{\kms}{$\,\mbox{km}\,\mbox{s}^{-1}$}
\newcommand{\ea}{\emph{et al.\ }}

\newcommand{\ha}{H$\alpha$ {}}
\newcommand{\lograt}{log (N$_{\rm Lyc}$/FUV)~}
\newcommand{\msol}{M$_{\odot}$}

\slugcomment{to be submitted to AJ, draft version: \today}

\shorttitle{Stewart \& Walter}
\shortauthors{UV--observations of the supergiant shell in IC\,2574}

\begin{document}

\title{UV Observations of the Powering Source of the Supergiant Shell in
IC\,2574}

\author{Susan G.\ Stewart}
\affil{U. S. Naval Observatory, 3450 Massachusetts Ave NW,  
Washington, DC 20392-5420}
\email{sgs@aa.usno.navy.mil }

\and

\author{Fabian Walter}
\affil{California Institute of Technology, Astronomy Dept. 105--24,
Pasadena 91125}
\email{fw@astro.caltech.edu}

\begin{abstract}
A multi-band analysis of the region containing the supergiant \ion{H}{1} shell
in the nearby dwarf irregular galaxy IC\,2574 presents
evidence of a causal relationship between a central star cluster, the
surrounding expanding \ion{H}{1} shell, and secondary star formation sites on
the rim of the \ion{H}{1} shell. Comparisons of the far--UV (FUV, 1521 \AA),
optical broad-band, H$\alpha$, X-ray, and \ion{H}{1} morphologies suggest that
the region is in an auspicious moment of star formation triggered by
the central stellar cluster. The derived properties of the \ion{H}{1} shell,
the central stellar cluster, and the star forming regions on the rim
support this scenario: The kinematic age of the \ion{H}{1} shell is $<$14
Myr and in agreement with the age of the central stellar cluster
derived from the FUV observations ($\sim$ 11 Myr). An estimate for the
mechanical energy input from SN and stellar winds of the
central stellar cluster made from FUV photometry and the derived
cluster age is $4.1 \times 10^{52}$ erg, roughly a few times higher than
the kinetic energy of the \ion{H}{1} shell. The requisite energy input needed
to create the \ion{H}{1} shell, derived in the `standard' fashion from the HI
observations (using the numerical models of Chevalier), is $2.6 \times
10^{53}$ erg which is almost an order of magnitude higher than the estimated
energy input as derived from the FUV data.  Given the overwhelming
observational evidence that the central cluster is responsible for the
expanding \ion{H}{1} shell, this discrepancy suggests that the required energy
input is overestimated using the `standard' method. This may explain
why some other searches for remnant stellar clusters in giant \ion{H}{1} holes
have been unsuccessful so far. Our observations also show that stellar
clusters are indeed able to create supergiant \ion{H}{1} shells, even at large
galactocentric radii, a scenario which has recently been questioned by
a number of authors.

\end{abstract}


\section{Introduction}

High--resolution observations in the 21\,cm line of neutral hydrogen
(HI) show that the interstellar medium (ISM) of galaxies is shaped in
a very complex way by holes and shells, most of which are expanding
(M\,31: Brinks \& Bajaja 1986, M\,33: Deul \& den Hartog 1990,
Holmberg I: Ott \ea 2000, Holmberg II: Puche \ea 1992, IC 10: Wilcots
\& Miller 1998, IC 2574: Walter \& Brinks 1999, DDO 47: Walter \&
Brinks 2000). Somewhat arbitrarily, these structures are coined `bubbles'
(diameters $\approx 10$\,pc), `superbubbles' ($\approx 200$\,pc) or
`supergiant shells' ($\approx 1000$\,pc, hereafter abbreviated SGSs).
In the standard picture (e.g., Weaver \ea 1977, McKee \& Ostriker
1977, Chu \ea 1995), these structures are believed to be created by
young star forming regions which supposedly eject a great amount
of mechanical energy into the ambient ISM in terms of strong stellar
winds and subsequent supernova (SN) explosions. In this picture, the
energy input of the massive stars creates a cavity filled with hot
ionized gas around the star forming region -- this overpressure drives the
expansion of a shell which collects the ambient neutral material on
its rim. For reviews on this topic, the reader is referred to
Tenorio--Tagle \& Bodenheimer (1988), van der Hulst (1996), Walterbos
\& Braun (1996), or Brinks \& Walter (1998).

Dwarf galaxies have proven to be ideal targets to study the largest of
these structures, the SGSs, because dwarf galaxies have puffed--up HI
disks, show almost solid body rotation and don't possess spiral
density waves. As a result, SGSs can form more easily and are not
destroyed prematurely, e.g., due to shear. We recently reported the
discovery of a particularly interesting SGS in IC\,2574 (Walter \ea
1998) which is the target of this study.

Although the standard picture to create these \ion{H}{1} structures (SN and
strong stellar winds) certainly sounds appealing, observational
evidence for this formation process is surprisingly scarce: only in few
cases are massive O and B type stars visible in bubbles or
superbubbles (see, e.g., N\,11: Mac~Low \ea 1998, or N\,44: Kim \ea
1998, both situated in the Large Magellanic Cloud).  However, in
the case of superbubbles, the ages derived for the stellar populations
often do not agree with the kinematic ages derived from the expansion
velocities of shell structures.  For example, from detailed surveys of
the massive stars in LMC superbubbles, Oey (1996) found `high velocity
superbubbles' containing central stellar associations with ages much
larger than the model kinematic ages.

Also, searches for the remnant stellar associations near the centers
of \ion{H}{1} cavities have not been particularly successful in the past --
the standard picture is therefore not without its critics. E.g., Rhode
\ea (1999) couldn't detect the expected number of remnant A and F
stars (as derived from the \ion{H}{1} observations) of the clusters which
supposedly created the expanding \ion{H}{1} holes in Holmberg~II. However,
Stewart \ea (2000) detect OB stellar populations interior to
several \ion{H}{1} holes in Holmberg II using ultraviolet data, a direct
tracer of the massive stellar component.

Another frequently used argument against creation of SGSs by
stellar associations is the location of some SGSs at large
galactocentric radii of a galaxy (where star formation is not expected
to play a dominant role).  Also, the largest cavities found seem to
surpass reasonable energy estimates based on single stellar clusters.

Since bubbles/superbubbles are often only detected in one or two
wavelengths it is difficult if not impossible to make meaningful
comparisons with theoretical models. Therefore, in depth
multi--wavelength studies are needed. Here, we focus on the SGS in
IC\,2574 (\S 2) and present evidence that this particular SGS has been
created by a massive stellar cluster in agreement with the standard
picture. This scenario is supported by new far ultraviolet (FUV)
observations of this region (\S 3), which allow us to independently
derive physical properties such as energy release and age of the powering 
stellar cluster of a SGS (\S 4). Some surprising results which also affect 
the interpretation of other SGSs will be discussed in
\S 5. The results and their implications are summarized in \S 6.

\section{The case of IC\,2574--SGS}

We aim to shed new insight on the controversy whether or not stellar
clusters can create SGSs by presenting new observations of an
expanding supergiant shell in the nearby dwarf galaxy IC\,2574
(hereafter referred to as IC\,2574--SGS, Walter \ea 1998).  The
\ion{H}{1} shell has a linear size of about 1000~pc $\times$ 500~pc
($\sim 60''\times30''$) and is expanding at $\approx$25~\kms. It is
therefore an ideal target to study expansion models since despite its
size it has not stalled yet (as most of the SGSs in other dwarf
galaxies have). The elliptical shape of the \ion{H}{1} shell in IC\,2574
is indicated in Fig.~1. The kinematic age ($t=r/v_{\rm exp}$) based on
the observed size and expansion velocity is estimated at 14 Myr. Note
that this age is actually an upper limit since the shell was
presumably expanding faster in the past. This indicates that the least
massive stars that go off as SN are most probably still present in the
central stellar association since their lifetimes ($\sim$ 50 Myr) are
somewhat longer then the dynamical age of the hole ($\sim$ 14 Myr, as
derived from the \ion{H}{1} observations).

Based on our \ion{H}{1} observations and using the models of Chevalier (1974),
we derive that the energy required to produce the shell must be of
order $10^{53}$ erg or the equivalent of about 100 Type II SN
(Walter \ea 1998). Note that in general, all time and shell
formation energy requirements are calculated from \ion{H}{1} shell size and
expansion velocity (e.g., using Chevalier's equation, see the
discussion in Walter \& Brinks 1999). However, before we begin
interpreting these energies, detailed studies of individual shells
must be conducted to insure these calculations make sense. The
available high resolution \ion{H}{1} data as well as the presence of the
interior stellar association renders IC\,2574--SGS an ideal target to
discern if this commonly used approach is indeed justified.

To do so, we present our analysis of FUV data of IC\,2574--SGS which
{\em for the first time gives us an independent measure} of the
age and the total mechanical energy deposited by the central stellar
association in a SGS.  These numbers can then be directly compared to
the energy derived from the \ion{H}{1} observations.

\section{The Observations}

\subsection{FUV observations}

The FUV observations were obtained by the Ultraviolet Imaging
Telescope (UIT) during the ASTRO-2 mission using the broad-band UIT B1
filter ($\lambda_{eff} = 1521$ \AA\ and $\Delta\lambda$ = 354
\AA). The IC\,2574 FUV image has 623 second exposure time and spatial
resolution of roughly $3\arcsec$. The image was calibrated using IUE
spectrophotometry of stars observed by the UIT. The FUV magnitudes are
computed from FUV flux using $m_{FUV}$= $-2.5\times
log_{10}(F(FUV)) - 21.1$, where F(FUV) is the incident flux in erg
(cm$^{2}$\AA~s) $^{-1}$. The chief photometric uncertainties are in
the form of low-level nonuniformities introduced via the development
and digitization processes. Uncertainty in the absolute calibration
can lead to uncertainties of up to 10 $-$ 15 \% in the FUV flux.  A
detailed discussion of the reduction of the UIT data to
flux-calibrated arrays is given by Stecher \ea (1997).

A minimal FUV background level ($\mu >$ 25 mag arcsec$^{-2}$) of
0.53$\pm0.1$ analog data units (ADU's) is computed by taking the mean
of 20, 20$\times$20 pixel boxes in areas void of any galaxy or stellar
flux. The UIT has limited accuracy at the lowest light levels. The
errors in the background levels reflect this intrinsic uncertainty.  A
wide, vertical stripe appeared in the center of a sizable percentage
of the UIT ASTRO-2 images, including the IC\,2574 image. The flaw
appeared on the film and was not a result of the digitization process;
its origin is unknown. It has been removed by a model which has been
constructed in parts of the image void of the galaxy.

The FUV image presented in Fig.~2 (left) shows a slight spiral
signature suggesting new star formation is preferential to the
arm-like regions. The enormous complex to the northeast containing
IC\,2574--SGS (as indicated on the figure) dominates the FUV morphology
of the galaxy. This complex alone contributes $\sim$50\% of the total
integrated FUV flux of IC\,2574.

\subsection{Optical observations}

CCD observations of IC\,2574 were obtained in the Johnson U and B bands,
and the Kron-Cousins R and \ha bands using the 1.0 m telescope at
Mount Laguna Observatory. The images are median combined with at
least two other images having the same exposure time to remove cosmic
rays and other defects. Astrometry and photometry are implemented
using standard IDL procedures for data reduction. The images are
calibrated using published photometry of seven stars in the galaxy
field (Sandage \& Tammann 1974).

The \ha filter at MLO has a centroid wavelength 6573 \AA\ with a width of
61~\AA\, which includes the \ion{N}{2} lines.  The \ha image is calibrated
using fluxes of 20 clearly defined \ion{H}{2}\ regions provided by
Miller \& Hodge (1994). An \ha emission image is produced by scaling
stars in the \ha image with those in the R--band image and subtracting
the stellar continuum component.

\subsection{FUV, optical and \ion{H}{1} morphology}

The relative FUV, HI, and \ha morphologies of the region containing
IC\,2574--SGS are illustrated in Fig.~2 (upper right). The FUV emission
is coincident with regions corresponding to both the interior of the
\ion{H}{1} shell and its rim.  The central stellar cluster (Fig.~2,
lower right) combined with star formation on the shell boundary
suggest the presence of secondary star formation triggered by the
expanding \ion{H}{1} shell (see, e.g., the models by Elmegreen 1994).
Note that \ha emission corresponds to regions on the rim of
IC\,2574--SGS while no \ha emission is present in its interior.  The 
FUV--band traces flux from
an evolving cluster for longer timescales (roughly 100 Myr) compared
to the \ha (roughly a few Myr). Figure~2 therefore already indicates
that the central cluster is relatively older than the star formation
on the shell boundary (see also \S~4.1).

This situation as well as the presumed presence of hot X--ray gas in
the center of the SGS (Walter \ea 1998) makes IC\,2574--SGS a
truly unique region and suggests that we have caught this SGS in an
auspicious moment. The morphology of the various wavelengths discussed
above suggest that this central stellar association is the powering
source for the formation and expansion of the shell as well as for the
heating of the X--ray gas. However, this evidence is yet only
circumstantial. Therefore, we present some physical properties of the
cluster/shell in the following analysis.

\section{Analysis of the data}

\subsection{Age analysis of IC\,2574--SGS}

We estimate the ages of the single regions on the rim of IC\,2574--SGS
as well as the central stellar association from their FUV, B, and \ha
fluxes. A similar approach is used to characterize the ages of star
forming regions in dwarf irregular Holmberg II by Stewart \ea (2000)
where it is described in full detail.  Circular apertures are used to
identify areas of associated FUV and \ha flux with the aid of a FUV,
\ha difference image, shown in Fig.~3. The twelve regions are defined
by selecting the circular aperture enclosing as much of the FUV and
\ha emission from a single region as possible.  This rather loose
criteria obviously neglects to include all the associated flux from a
single region and has the potential to mix regions of different
populations, but it is sufficient to discern the relative ages of star
formation regions (see Stewart \ea and references therein).

Listed in Table 1 are the blue magnitude (m$_{\rm B}$), the H$\alpha$ flux, the FUV
flux, and the FUV magnitude (m$_{\rm FUV}$) for each region.  Values
are corrected for Galactic foreground extinction.  Also given in 
Table 1 are the FUV$-$B color (m$_{\rm FUV}$ -- m$_{\rm B}$) and 
log (N$_{\rm Lyc}$/FUV). The latter quantity is the logarithmic ratio
of the number of Lyman continuum photons, converted from \ha flux
assuming Case B recombination, to FUV flux (where $N_{\rm Lyc}/FUV =
(F(H\alpha)/3.02 \times 10^{-12} \times 0.453)/F(FUV)$). The FUV$-$B
color and \lograt are time dependent quantities which vary over the
lifetime of a cluster and can be compared to model values to estimate
an age from the observables.

A single generation instantaneous burst (IB) model which employs the
stellar evolutionary tracks of Schaerer \ea (1993) and the stellar
atmosphere models of Kurucz (1992) is used to derive the expected flux
from an evolving cluster in an environment similar to that of
IC\,2574. The model estimates L$_{\rm FUV}$, N$_{\rm Lyc}$, and L$_{\rm
B}$ assuming an initial mass function (IMF) and metallicity. Figures 4
and 5 show the time dependence of the FUV$-$B color and \lograt of a
model cluster. To illustrate the model dependence on these assumed
parameters, we calculate various combinations of IMF and metallicity.
Figure 6 shows the time evolution of \lograt vs.\ FUV$-$B color for a
cluster based on the results of Figures 4 and 5. The figure depicts
the model assuming a Salpeter IMF (2.35) and an SMC-like metallicity,
Z/Z$_{\odot}=0.1$ (as derived for IC\,2574 by Miller \& Hodge 1996),
which is used in the subsequent derivations. Here, each point
represents one age (steps: 0.5\,Myr), starting with a 0.5\,Myr old
cluster in the upper left-hand corner of the graph. The last point on the
lower right represents a cluster of an age of 50\,Myr.

We use this plot to compare the theoretical model with the actual
observed values for FUV$-$B and log (N$_{\rm Lyc}$/FUV). Since both axes represent
time-dependent quantities, the relationship between the observed value
and the model is indicative of the age. 

Note that most of the observed data points lie off the model since
they are uncorrected for internal extinction. Region~1, however (the
central stellar cluster; x$_1$ on the graph) is not far off the
theoretical curve. This is not really surprising since we don't
expect redenning to play a huge role in this case (since the cluster
is situated within the \ion{H}{1} cavity). The other regions are
obviously still embedded in their parental molecular clouds, which
nicely explains the higher redenning.

In order to make an age estimate, the observed FUV$-$B color and \lograt are
corrected for internal extinction effects.  The assumed redenning
vector is indicated in the upper right of Fig.~6.  The SMC redenning
curve is used under the assumption that the shape of the curve
correlates with metallicity (Verter \& Rickard 1998).  In the FUV,
A$_{\rm FUV}$/E(B$-$V)=17.72 (Hill \ea 1997). Each data point is
corrected until the point at which it agrees with the theoretical
curve. The amount of this correction is an estimate of the internal
extinction; the position on the theoretical curve after the correction
yields the age of the cluster.  The age obtained using this method
should be treated as the mean age of stars in the aperture since a
single generation model is used to interpret flux from what is
potentially a mix of populations of slightly different ages.  The age
estimates are given for each region in Table~2 (column~1).
 
\subsection{Energy analysis}

The observed FUV flux is used to estimate the mechanical energy
imparted to the surrounding ISM of each cluster over its lifetime. The
FUV luminosity, L$_{\rm FUV}$, of each region is calculated from the
FUV flux after correction for the internal extinction (derived in \S
4.1), assuming a distance of 3.2 Mpc (Table~2, column~2). The mass of
each region is then estimated by comparing the observed L$_{\rm FUV}$
with the model L$_{\rm FUV}$ (erg s$^{-1}$ ~\msol$^{-1}$) at the
cluster's estimated age. The derived masses are given in Table~2
(column~3).

Evolutionary synthesis models of populations of massive stars given by
Leitherer \& Heckman (1995, Fig. 55) provide estimates of the
deposition rates, L$_{\rm mech}$, from stellar winds and SN
for an instantaneous burst as a function of time and mass. An average
rate of L$_{\rm mech}\sim2\times10^{34}$ erg (s~ \msol)$^{-1}$ is
derived for clusters of similar IMF and metallicity from the
models. An estimate of the net mechanical energy deposited into the
ISM by SN and stellar winds, E$_{\rm mech}$, can be made by
taking the product of L$_{\rm mech}$, the cluster mass and the cluster
age ($E_{\rm mech} \sim L_{\rm mech} \times {\mbox mass} \times {\mbox
age}$).  The derived values of L$_{\rm FUV}$ and E$_{\rm mech}$ are
given for each cluster in Table~2 (columns 4 and 5, respectively).

\section{Discussion}

The analysis of the ages of star forming regions indicates a dichotomy
between the age of the central cluster (region1) and that of the
surrounding regions (regions 2--12). The derived age for the central
cluster is $\sim$11 Myr while the other regions (which all coincide
with the rim of the \ion{H}{1} shell) range in age from $\sim$1$-$4.5
Myr (average age: 3.1 Myr). This is also apparent in Fig.~6 where the
data point for region1 is found at the lower end of the plot, set off
from the rest of the regions.  Like the comparison between the FUV and
\ha morphologies in Fig.~2, our age analysis suggests sequential
star formation on the rim of the \ion{H}{1} shell, triggered by the
expansion of the shell. The age of the central cluster agrees very
well with the upper limit derived independently from the \ion{H}{1}
observations ($\sim$14 Myr). This provides strong independent evidence
that the cluster indeed created IC\,2574--SGS.

The mechanical energy provided over the lifetime of the central
cluster is estimated to be E$_{\rm mech}\sim 4.1 \pm 0.8
\times 10^{52}$ erg, roughly a factor of two times the kinetic energy of
the
expanding shell as derived from the \ion{H}{1} data ($1.7 \pm 0.5
\times 10^{52}$, Walter \ea 1998). It is now important to 
compare the total energy estimate based on the FUV observations with
previous estimates based on the \ion{H}{1} data
only. \ion{H}{1}--based estimates are usually performed using
Chevalier's equation. In the case of IC\,2574--SGS, such an energy
estimate is 6 times higher than the FUV value ($2.6 \pm 1.0 \times
10^{53}$\,erg, Walter \ea 1998). It is not really surprising that
these values do not agree better since the model by Chevalier is
based on the late phase of SN remnants and simply scaled
up to supergiant shells. Our result therefore seems to indicate that
\ion{H}{1} based energy estimates using Chevalier's equation
overestimate the actual energy needed to create an \ion{H}{1}
shell. This has important consequences for searches for remnant
stellar clusters in other SGSs which are based on an expected
luminosity of the cluster as derived from the \ion{H}{1} data (e.g.,
Rhode \ea 1999). In other words, if our result holds true for other
SGSs as well, one may overestimate the luminosity of a remnant stellar
cluster by almost an order of magnitude.

\section{Summary and Conclusion}

A multi-band analysis of the region containing the supergiant \ion{H}{1} shell
in the dwarf galaxy IC\,2574 presents evidence of a causal
relationship between a central star cluster, a surrounding expanding HI
shell, and secondary star formation sites on the rim of the shell.

The key results from the study can be summarized as follows:

(1) A stellar cluster interior to the expanding supergiant \ion{H}{1} shell in IC\,2574 is identified in the FUV and optical bands.

(2) The center of the \ion{H}{1} shell is void of H$\alpha$ emission while
\ion{H}{2}\ regions coincide with the shell boundary. This is
confirmed by the FUV, \ha difference images indicating propagating
star formation outwards from the center of the shell.

(3) A detailed age analysis of the star forming regions reveals that the
central cluster is $\sim$11 Myr and the oldest in the complex. This
age agrees very well with the value derived independently from the HI
observations (upper limit of $\sim$14 Myr).

(4) Analysis of the FUV flux in the central star forming region
indicates that the mechanical energy imparted to the ISM from
SN and stellar winds over the lifetime of the cluster is $\sim
4.1 \pm 0.8 \times 10^{52}$ erg, roughly a factor of two greater than
the kinetic energy of the expanding shell.

(5) The numerical models of Chevalier suggest an energy input of $2.6
\pm 1.0 \times 10^{53}$ erg needed to create the \ion{H}{1} shell,
roughly a factor of six greater than the amount of energy input
derived using the FUV observations. The fact that the classical method
(using the models of Chevalier) seems to overestimate the required
energy by almost an order of magnitude suggests caution in using it as
the basis for determining the required energy input for \ion{H}{1}
shells. This has important consequences for searches of remnant
stellar clusters based on the \ion{H}{1} data alone.

(6) Our results support the `standard' picture of the creation of
supergiant shells by massive star clusters (even at large
galactocentric distance). We conclude that a combination of SN
explosions and strong stellar winds are most likely responsible for
the supergiant shell in IC\,2574.

Our results show that indeed massive stellar clusters can create
supergiant shells in galaxies (even at large galactocentric distance)
as predicted by the `standard' picture (creation by SN explosions and
strong stellar winds). One intriguing question remains, of course: why
don't we see similar stellar clusters in other supergiant \ion{H}{1}
shells in IC\,2574 as well as in other galaxies? One answer might be
that in the case of IC\,2574--SGS we were just lucky enough to have
caught this particular shell in an auspicious moment. We speculate
that we may have difficulties to detect the central cluster in
IC\,2574 after some $\sim 10^8$ years (a typical age for the largest
\ion{H}{1} structures found in other galaxies) both because of dimming
and dispersion of the cluster stars (since the cluster may not be
gravitationally stable after all the gas has been blown away). This
would mean that chances to detect similar structures in other galaxies
are low, making IC\,2574--SGS a truly unique object to study the
evolution of supergiant shells in general.

\acknowledgements

FW acknowledges NSF grant AST 9613717.

\clearpage

\begin{figure}
\caption{left: \ion{H}{1} image of IC\,2574 with IC\,2574--SGS marked; 
right: \ha image of the central cluster with an ellipse marking the
position of the \ion{H}{1} shell boundary. The arrow indicates the
North direction.}

\caption{left: FUV image of IC\,2574 (same region as Fig.~1, left); 
upper-right: FUV image of the central cluster with an ellipse marking
the \ion{H}{1} shell boundary (same region as Fig.~1, right). The
contours represent 5, 15, 30, and 50 \% of the peak \ha brightness;
lower-right: the R-band image of the central cluster with an ellipse
marking the \ion{H}{1} shell boundary. The orientation is the same as
in Fig.~1.}

\caption{FUV, \ha difference image with apertures indicating the regions 
defined by us. The image is displayed so that areas containing FUV and
no \ha are black and areas containing \ha and no FUV are white. The
orientation is the same as in Fig.~1.}

\caption{Plot illustrating the age dependence of FUV--B color for a 
single generation instantaneous burst cluster model.  The lines
represent a model of LMC metallicity (Z/Z$_{\odot}=0.4$) and IMF of
1.08 (solid line); a SMC metallicity (Z/Z$_{\odot}=0.1$) and IMF of
1.08 (dash-dotted line); a LMC metallicity and IMF of 2.35 (dotted
line); and a SMC metallicity and IMF of 2.35 (solid line with filled
circles). }

\caption{Plot illustrating the age dependence of the  
logarithmic ratio of the number of Lyman continuum photons to FUV flux
for a single generation instantaneous burst cluster model. The lines
are the same as for Fig.~3.}

\caption{The plot of model \lograt and FUV--B assuming SMC metallicity and 
Salpeter IMF (based on Fig.~4 and~5). The graph is for a 50 Myr
cluster, with each dot representing 0.5 Myr (beginning with a 0.5 Myr
old cluster in the upper left-hand corner of the graph). The data
points represent the observed values for each region before correction
for internal extinction. The subscripts on the data points represent
the region numbers ($\it i.e.$ x$_1$ = region 1). }
 
\end{figure}
\clearpage

%
%
\begin{table*}
\begin{center}
\begin{tabular}{ccccccc}
\tableline\tableline
Region   &m$_{\rm B}$  & F(H$\alpha$)   &F(FUV) &m$_{\rm FUV}$ &FUV$-$B  & \lograt \\
   \#    &             & 10$^{-15}$ erg (cm$^{2}$\AA~s)$^{-1}$ &10$^{-15}$ erg (cm$^{2}$\AA~s)$^{-1}$&  &  &  \\
\tableline
1       & 16.23  & 13  &  19.0    &       13.20           & $-$3.03 & 11.69\\
2       & 17.09  & 445 &   5.4    &       14.57           & $-$2.52 & 13.78\\
3       & 20.25  & 24  &   0.8    &       16.61           & $-$3.64 & 13.33\\
4       & 16.79  & 296 &  11.6    &       13.74           & $-$3.05 & 13.27\\
5       & 18.17  & 190 &   3.2    &       15.15           & $-$3.02 & 13.64\\
6       & 15.68  & 1000&  31.5    &       12.65           & $-$3.03 & 13.36\\
7       & 18.80  & 79  &   2.5    &       15.42           & $-$3.38 & 13.37\\
8       & 17.83  & 62  &   4.8    &       14.70           & $-$3.13 & 12.98\\
9       & 19.76  & 71  &   0.9    &       16.46           & $-$3.30 & 13.74\\
10      & 19.68  & 71  &   0.7    &       16.72           & $-$2.96 & 13.84\\
11      & 17.61  & 120 &   6.0    &       14.45           & $-$3.16 & 13.16\\
12      & 19.32  & 55  &   1.2    &       16.22           & $-$3.10 & 13.53\\     
\tableline
\end{tabular}
\end{center}
\tablenum{1}
\caption{Observations for Star Forming Regions in IC\,2574--SGS \label{Table 1}}
\end{table*}
\clearpage

%
\begin{table*}
\begin{center}
\begin{tabular}{lcccr}
\tableline\tableline
Region     &  Age         & L$_{\rm FUV}$  & Mass    &  E$_{\rm mech}$  \\
   \#      & (Myr)        & (10$^{36}$ erg s$^{-1}$)& $10^3$\msol & (10$^{50}$erg)\\
\tableline
1  &  25.3       &  25.3     & 142 &412.0     \\
2  &  33.7       &  33.7     & 107 &84.5      \\
3  &   1.3       &   1.3     &   4.5 &3.0     \\
4  &  29.7       &  29.7     & 113 &121.4     \\
5  &  10.3       &  10.3     &  36 &23.8      \\
6  &  94.9       &  94.9     & 308 &316.1     \\
7  &  5.3        &  5.3      &  16 &14.0     \\
8  &  8.1        &  8.1      &  27 &32.3      \\
9  &  2.1        &  2.1      &   7.8 &4.1      \\
10 &  2.4        &  2.4      &  10 &2.7      \\
11 &  12.0       &  12.0     &  39 &43.2       \\
12 &   3.5       &   3.5     &  11 &8.9     \\
\tableline
     
\tablecomments{The FUV luminosity is
derived from flux corrected for internal extinction and an assumed
distance of 3.2 Mpc.}
\end{tabular}
\end{center}
\tablenum{2}
\caption{Derived Parameters for Star Forming Regions in IC\,2574--SGS
\label{Table 2}}
\end{table*}

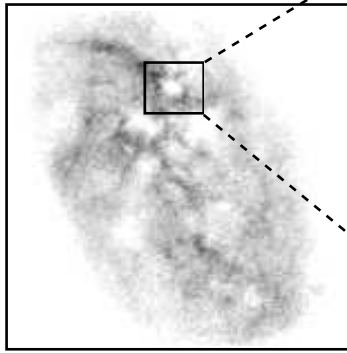
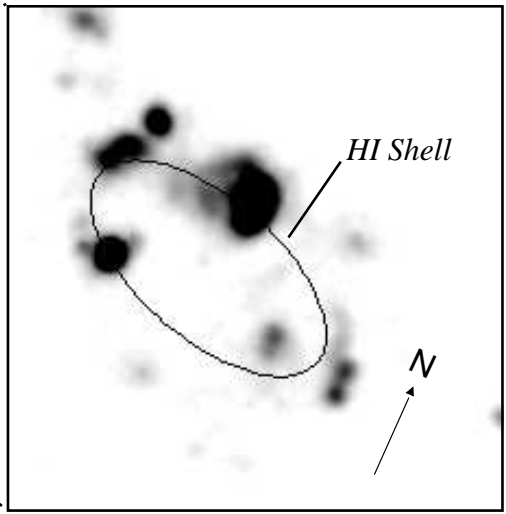
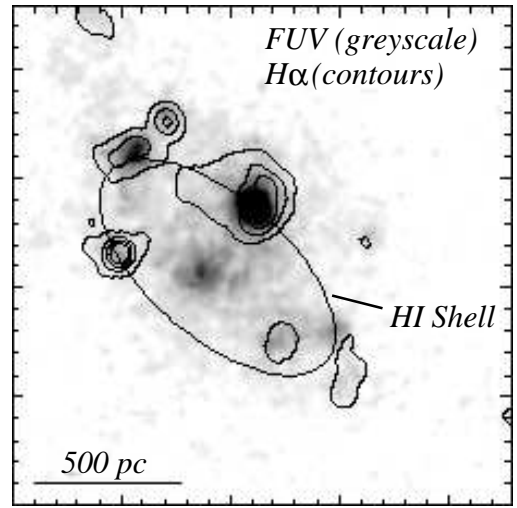
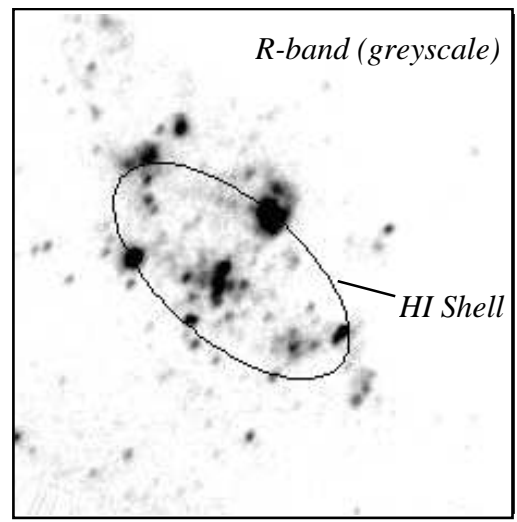
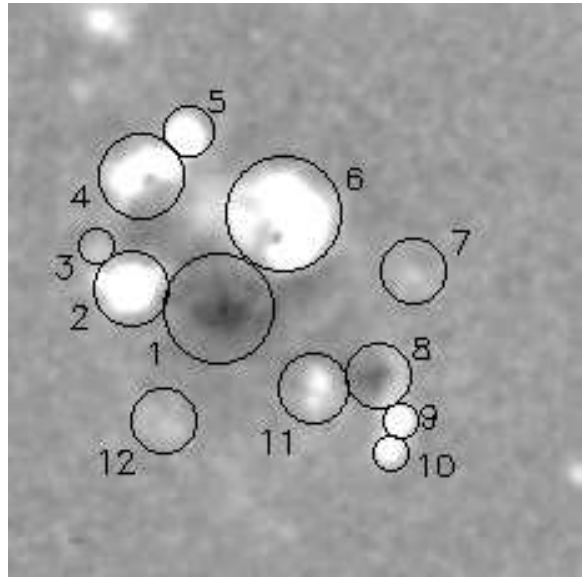
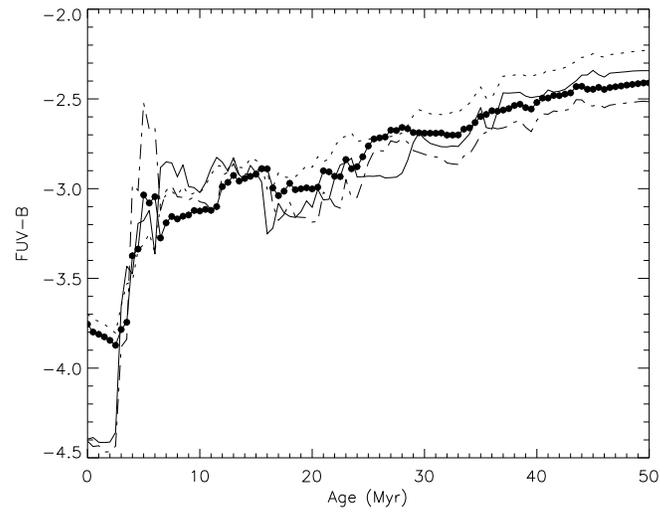
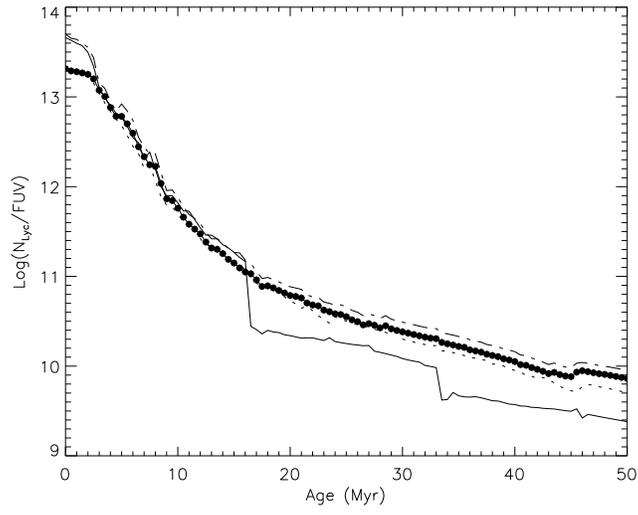
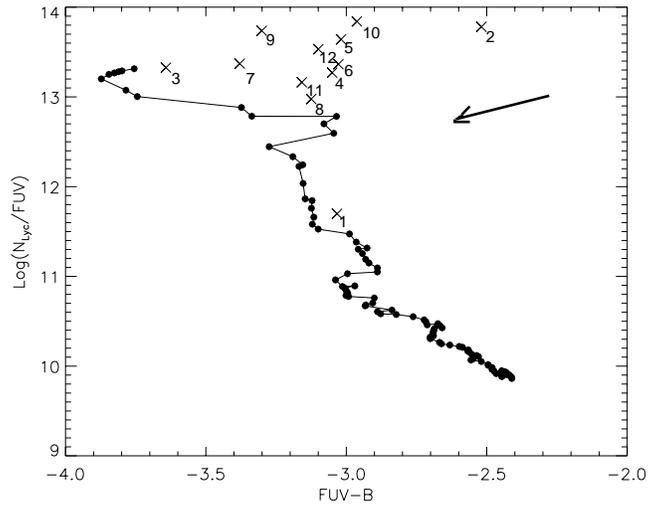
\end{document}